\documentclass[prl,aps,twocolumn,showpacs]{revtex4}

\usepackage{epsfig}

\begin{document}
\title{Density scaling as a property of strongly correlating viscous liquids}
\author{Thomas B. Schr{\o}der, Ulf R. Pedersen, and Jeppe C. Dyre}
\affiliation{DNRF Centre ``Glass and Time,'' IMFUFA, Department of Sciences, Roskilde University, Postbox 260, DK-4000 Roskilde, Denmark}
\date{\today}

\begin{abstract}
We address a recent conjecture according to which the relaxation time $\tau$ of a viscous liquid obeys density scaling ($\tau=F(\rho^\gamma/T)$ where $\rho$ is density) if the liquid is ``strongly correlating,'' i.e., has almost 100\% correlation between equilibrium  virial and potential-energy fluctuations [Pedersen {\it et al.}, PRL {\bf 100}, 011201 (2008)]. Computer simulations of two model liquids - an asymmetric dumbbell model and the Lewis-Wahnstr\"om OTP model - confirm the conjecture and demonstrate that the scaling exponent $\gamma$ can be accurately predicted from equilibrium fluctuations. 
\end{abstract}


\maketitle

Understanding the glass transition depends on understanding the preceding highly viscous liquid phase, where the relaxation time upon cooling approaches and exceeds seconds \cite{gt_rev}. Increasing the pressure also leads to much slower relaxations. The study of glass-forming liquids under high pressure has recently become popular, and many data are now available on the properties of the alpha and beta processes, etc., under pressure. If the density is $\rho$ and $T$ is temperature, the last few years have shown that for many highly viscous liquids the (alpha) relaxation time follows the scaling expression
\begin{equation}\label{1}
\tau\,=\,
%F\left(\frac{\rho^\gamma}{T}\right)\,.
F\left({\rho^\gamma}/{T}\right)\,.
\end{equation}
The state of this rapidly developing field as of 2005 was summarized in the review Ref. \cite{rol05} by Roland {\it et al.} that presented data for more than 50 liquids and polymers.

Equation (\ref{1}) defines what is referred to as thermodynamic or density scaling. Density scaling is a recent discovery that, following pioneering works by T{\"o}lle and Dreyfus {\it et al.} \cite{tol01,dre03}, was proposed as a general principle in 2004 in papers by Alba-Simionesco and co-workers and by Casalini and Roland \cite{alb04,cas04}. The former authors demonstrated data collapse at varying temperature and density following a more general expression than Eq. (\ref{1})  used by the latter authors.

Dreyfus and co-workers found $\gamma=4$ for ortho-terphenyl (OTP) and argued that this could be linked to the $r^{-12}$ repulsive term of the Lennard-Jones potential. It turned out, however, that $\gamma=4$ is not a special exponent \cite{rol05}, leaving the question of the microscopic interpretation of $\gamma$ open. Coslovich and Roland very recently addressed this by computer simulations of binary Lennard-Jones like liquids where the exponent of the repulsive part of the potential took the values $8,12,24,36$ \cite{cos08}. These model systems obey density scaling and the exponent $\gamma$ is to a good approximation 1/3 of the exponent characterizing the effective power law of the repulsive core of the potential, as expected for an exact inverse power-law potential \cite{cos08} (see also Ref.~\cite{mic04}). 

Recently, simulations of the thermal equilibrium fluctuations of pressure, energy, and volume in different ensembles revealed that these quantities correlate strongly for a number of model systems \cite{ped_pre, ped_prl}. For instance, in the NVT ensemble (i.e., constant volume and temperature) the following systems are ``strongly correlating'' in the sense that they show more than 90\% correlation between virial (the non-kinetic part of the pressure) and potential energy: The standard Lennard-Jones liquid, a liquid with exponential short-range repulsion, the Kob-Andersen binary Lennard-Jones liquid,  a seven-site united-atom model of toluene, and the model system studied below consisting of  asymmetric ``dumb-bell'' type molecules. The correlations derive from the repulsive core of the intermolecular potential that, interestingly, dominate fluctuations even at zero and slightly negative pressure \cite{ped_prl,nick}. For the standard Lennard-Jones liquid the repulsive core is approximately described by a repulsive $r^{-18}$ term \cite{ped_prl,ben03}. The exponent of the approximate power law depends weakly on state point. 

In view of the fact that both density scaling and the strong correlations reflects an effective inverse power-law of the repulsive core of the potential, the following conjecture was proposed in Ref. \cite{ped_prl}: A viscous liquid is strongly correlating if and only if it obeys density scaling to a good approximation. There is evidence indirectly supporting this: Highly viscous liquids with strong hydrogen bonds are not strongly correlating because of the ``competing interactions'' \cite{ped_prl}, and these liquids also do not obey density scaling very well \cite{rol05,grz06,rol08}. In the present publication simulations are presented that supports the conjecture and strengthens it by adding: {\it For strongly correlating viscous liquids density scaling is obeyed with a scaling  exponent $\gamma$ that can be determined from thermal equilibrium virial and potential energy fluctuations.} This amounts to the simplest possible assumption; that it is the same part of the potential that dominates equilibrium fluctuations and flow dynamics.

We performed NVT molecular dynamics simulations  \cite{SimDet} of 512 asymmetric dumbbell molecules consisting of pairs of Lennard-Jones (LJ) spheres connected by rigid bonds. The dumbbells were parameterized to mimic toluene \cite{Dumbbell}. Charges of $\pm q$ (specified below) were applied to the LJ spheres. The model was simulated with two charges: i) $q=0$ corresponding to the simulations done in  Refs. \cite{ped_pre} and \cite{ped_prl}. This version of the model is a ``strongly correlating viscous liquid''. ii) $q=0.5$e (e being the elementary charge) resulting in a dipole moment of %0.29nm$\cdot$0.5e=2.32$\cdot10^{-29}$Cm=
7.0D, i.e., almost 20 times stronger than in toluene,
%(0.36D)
 and almost 4 times stronger than water.
% (1.8D)
The purpose of using such a large value of $q$ was to break the correlations and thus to have a version of the model that is \emph{not} strongly correlating. 

\begin{figure}
\begin{center} 
 \includegraphics[width=8.5cm]{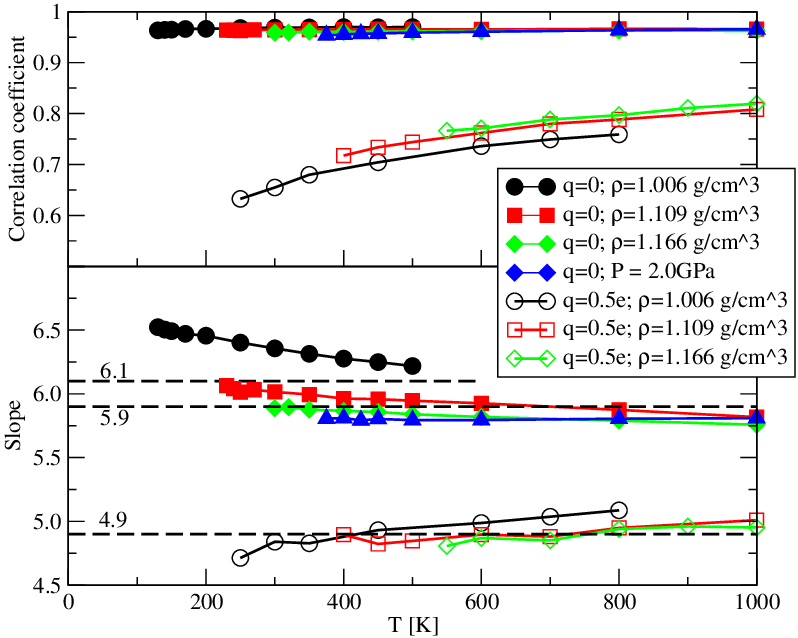}
\end{center}
\caption{
Results from equilibrium molecular dynamics simulations of 512 asymmetric dumbbell molecules with, respectively, a strong dipole moment ($q=0.5$e, open symbols, three isochores), and zero dipole moment ($q=0$e, filled symbols, three isochores and an isobar). (a) Correlation coefficients, 
%describing the correlation of fluctuations in virial, $W$, and potential energy, $U$: 
$R\equiv \left< \Delta W \Delta U\right>/\sqrt{ \left< \Delta W^2 \right>\left< \Delta U^2 \right>}$. (b) The 'slopes', $\gamma \equiv \sqrt{\left< (\Delta W)^2 \right>/\left< (\Delta U)^2 \right>}$.
}\label{Fig:CCSlope}
\end{figure}

If the virial is denoted by $W$, and $U$ is the potential energy, $\Delta W(t) \equiv W(t) - \left < W\right>$ and $\Delta U(t) \equiv U(t) - \left < U\right>$, where $\left < ... \right>$ indicates thermal average. The correlation coefficient is defined by $R\equiv \left< \Delta W \Delta U\right>/\sqrt{ \left< \Delta W^2 \right>\left< \Delta U^2 \right>}$. $R$ is plotted for a several state points in Fig. \ref{Fig:CCSlope}(a). For $q=0$ (filled symbols) all investigated state points have $R>0.95$. For $q=0.5$e (open symbols) the correlation coefficient is significantly smaller; the Coulomb interactions do indeed break the correlations as expected \cite{ped_prl}.

We define $\gamma \equiv \sqrt{\left< (\Delta W)^2 \right>/\left< (\Delta U)^2 \right>}$. If $R\approx 1$ it follows that $\Delta W(t) \approx \gamma \Delta U(t)$ in their instantaneous fluctuations \cite{ped_prl}, and consequently we refer to $\gamma$ as the 'slope'. According to our conjecture, the slope $\gamma$ is also the scaling exponent in Eq.~(1). In Fig.~\ref{Fig:CCSlope}(b) we show the slopes for all investigated state points. For $q=0$ (filled symbols) there is a small, but significant dependence on density and temperature. Thus if the conjecture is correct, density scaling can not be exact: To apply density scaling a single value of $\gamma$ is needed, but the $\gamma$ we get from the fluctuations depends slightly on the state point \cite{ped_prl}. In the following  we consider for $q=0$ two values of the scaling exponent: the slope averaged over all state points with $q=0$;  $\gamma = 6.1$, and the slope averaged over the three data sets with the smallest slopes for $q=0$ ($\rho = 1.109$ g/cm$^3$, $\rho = 1.166$ g/cm$^3$ and P=2.0GPa); $\gamma =5.9$, i.e., the 'best' compromise if we chose to ignore the $\rho = 1.006$ g/cm$^3$ isochore. For $q=0.5$e the slopes are less density dependent with a mean value $\gamma =4.9$.

\begin{figure}
 \begin{center}
 \includegraphics[width=8.5cm]{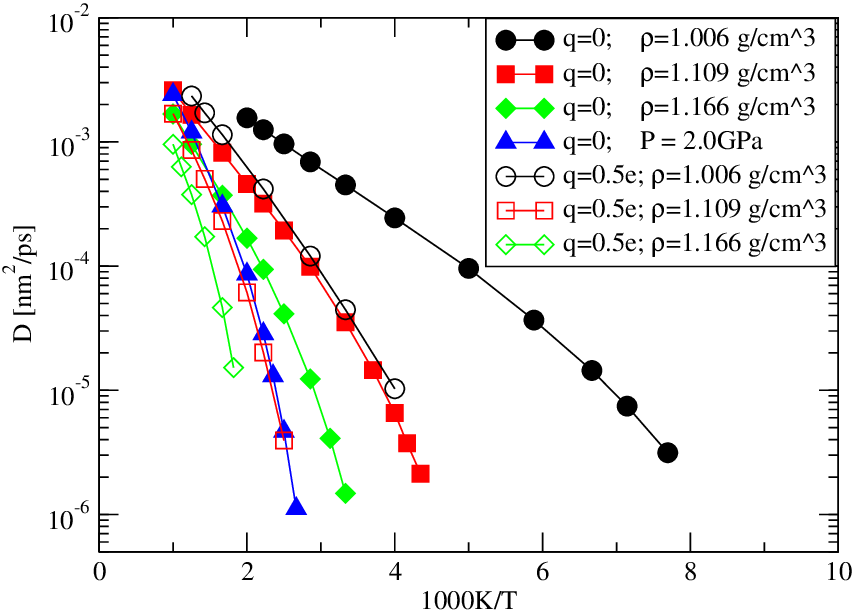}
\end{center}
\caption{Arrhenius plot of the diffusion coefficient, $D$, for the asymmetric dumbbell model.
}\label{Fig:D}
\end{figure}

In the following we apply density scaling to the diffusion coefficient estimated from the long-time behavior of the mean-square displacement, $\left< \Delta r^2(t) \right>$, of the large spheres (the ``phenyl group'') \cite{Equil}. The diffusion coefficients for all state points studied are given in Fig.~\ref{Fig:D}. 

\begin{figure}
\begin{center}
  \includegraphics[width=8.5cm]{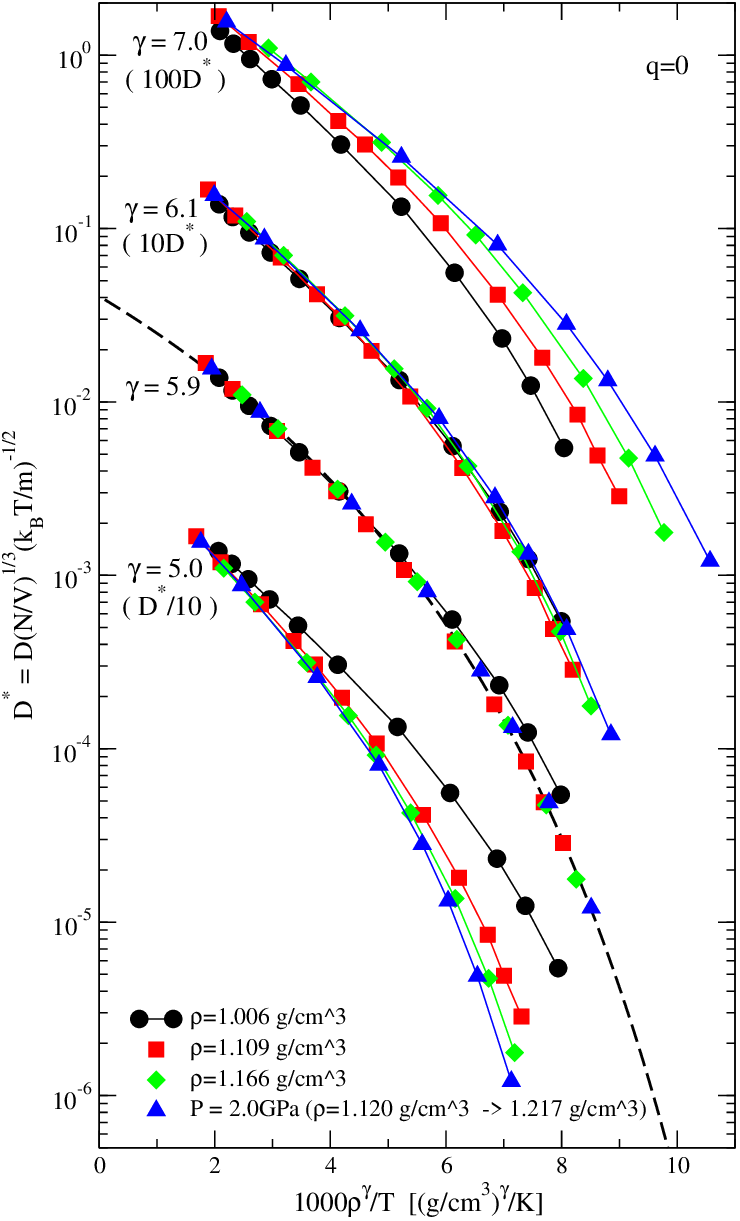}
\end{center}
\caption{
The reduced diffusion coefficient,  $D^* \equiv (N/V)^{1/3}(k_BT/m)^{-1/2}D$, for the asymmetric dumbbell model with $q=0$ scaled according to Eq.~(1), with four different scaling exponents:
$\gamma = 7.0$ (Upper set of curves, $D^*$ multiplied by 100), 
$\gamma = 6.1$ (Second set of curves, $D^*$ multiplied by 10), 
$\gamma = 5.9$ (Third set of curves), and 
$\gamma = 5.0$ (Lower set of curves, $D^*$ divided by 10).
As a guide to the eye, the equation 
$D^* = 4.07\times 10^{-2} \exp\left( -462/(T/\rho^{5.9} - 60.8)\right)$ is plotted as a 
fit to the three collapsing curves for $\gamma = 5.9$.
}\label{Fig:DScaled}
\end{figure}

Following  Coslovich and Roland \cite{cos08} we apply density scaling to the reduced diffusion coefficient, $D^* \equiv (N/V)^{1/3}(k_BT/m)^{-1/2}D$ where $m$ is the mass of the molecules. In Fig.~\ref{Fig:DScaled}(a) $D^*$ is plotted for $q=0$ as a function of $1000\rho^\gamma/T$ with four different values of $\gamma$. Clearly, density scaling works neither with $\gamma=7.0$ or $\gamma=5.0$. Comparing the scaling with $\gamma = 6.1$ to the data without scaling (filled symbols in Fig. \ref{Fig:D}), we find here good data collapse; by far most of the density dependence is captured by the density scaling with $\gamma = 6.1$.  With $\gamma = 5.9$ the data collapse is even better for three of the data sets, whereas one data set deviates slightly from the master curve comprised of these three sets. This is the isochore $\rho=1.006$ g/cm$^3$, i.e., the one that was ignored when choosing $\gamma = 5.9$ (Fig.~\ref{Fig:CCSlope}(b)).

The conclusion drawn from Fig.~\ref{Fig:DScaled} is two-fold: i) Density scaling is approximate (as discussed above) - for a larger region of state points scaling will be less perfect. ii) For a given range of state points, the scaling exponent can be found by studying equilibrium fluctuations.

\begin{figure}
\begin{center}
 \includegraphics[width=8.5cm]{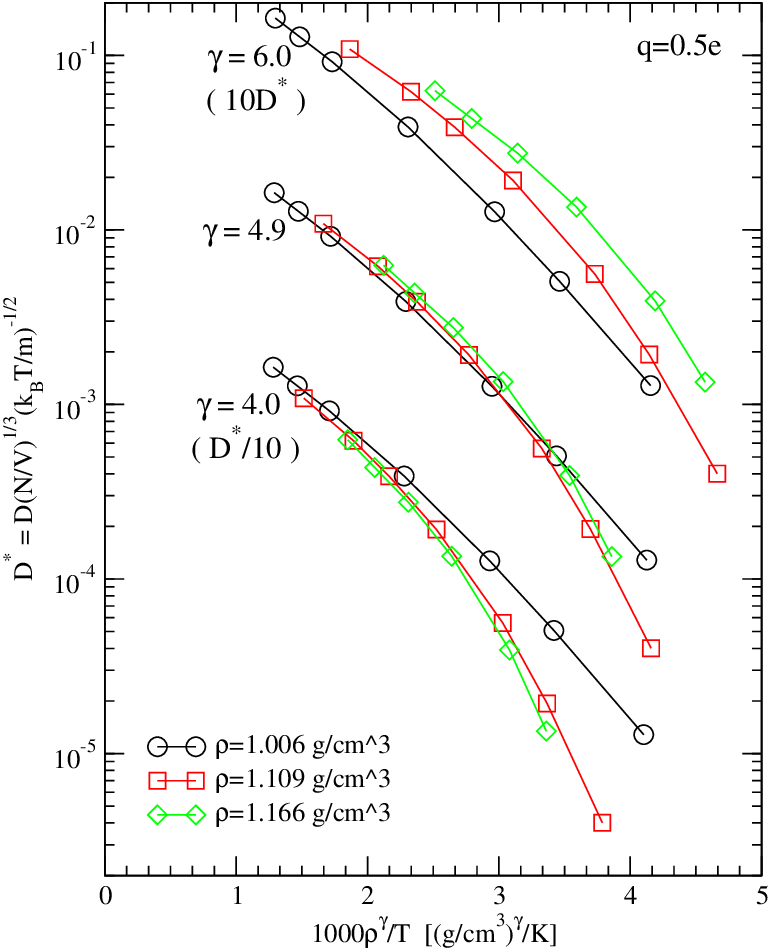}
\end{center}
\caption{
The reduced diffusion coefficient, $D^* \equiv (N/V)^{1/3}(k_BT/m)^{-1/2}D$, for the asymmetric dumbbell model with $q=0.5$e scaled according to Eq.~(1), with three different scaling exponents:
$\gamma = 6.0$ (Upper set of curves, $D^*$ multiplied by 10), 
$\gamma = 4.9$ (Middle set of curves), and 
$\gamma = 4.0$ (Lower set of curves, $D^*$ divided by 10).}
\label{Fig:DScaled_q}
\end{figure}

In Fig.~\ref{Fig:DScaled_q} the reduced diffusion coefficients $D^*$ for $q=0.5$e are plotted as a function of $1000\rho^\gamma/T$ with three different values of $\gamma$. The value $\gamma = 4.9$ chosen from the equilibrium fluctuations (Fig.~\ref{Fig:CCSlope}(b)), is found to be a reasonable scaling exponent. However, as conjectured, the data collapse achieved is inferior to that for the strongly correlating version of the model ($q=0$). 

\begin{figure}
\begin{center}
 \includegraphics[width=8.5cm]{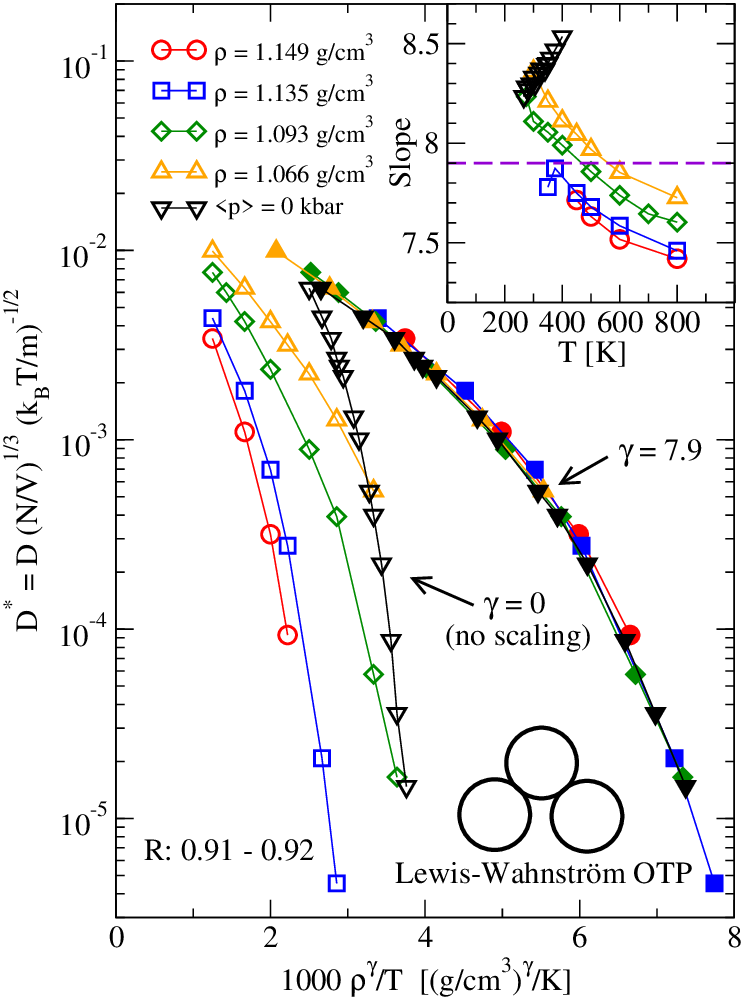}
\end{center}
\caption{
The reduced diffusion coefficient, $D^* \equiv (N/V)^{1/3}(k_BT/m)^{-1/2}D$, for LW-OTP \cite{OTP}. For the zero pressure isobar the densities covers the interval 1.008 g/cm$^3$ to  1.089 g/cm$^3$ Filled symbols: $D^*$ scaled according to Eq.~(1) with $\gamma=7.9$ chosen from the slopes (inset). Open symbols: $D^*$ plotted without scaling ($\gamma=0$).
}
\label{Fig:DScaled_OTP}
\end{figure}

To test the generality of our findings, we repeated the analysis for the Lewis-Wahnstr\"om OTP model (LW-OTP) \cite{SimDet,OTP} (see also Refs. \cite{cho04} and \cite{tar04}). The results achieved for this model (Fig.~\ref{Fig:DScaled_OTP}) are qualitatively similar to the results for the asymmetric dumbbell model with $q=0$; (i) LW-OTP is strongly correlating ($0.91<R<0.92$) \cite{OTP-R}, (ii) the slope is slightly state point dependent, (iii) choosing the average slope as scaling exponent $\gamma$ gives good data collapse.

In summary, we have presented numerical evidence for the conjecture that density scaling is a property of strongly correlating viscous liquids. For the two strongly correlating models investigated density scaling applies with a scaling exponent that can be accurately predicted from the equilibrium fluctuations. This represents a  step forward in the theoretical understanding of density scaling. In particular, the scaling exponent $\gamma$ should no longer be regarded as an empirical fitting parameter. In computer simulations $\gamma$ can be estimated directly from the equilibrium fluctuations. Via the fluctuation dissipation theorem the equilibrium fluctuations determine the frequency-dependent linear thermoviscoelastic response functions - this fact provides a possible future route for independent experimental estimation of the scaling exponent $\gamma$ \cite{ped_pre,ell07}.

\acknowledgments 
This work was supported by the Danish National Research Foundation's (DNRF) centre for viscous liquid dynamics ``Glass and Time.''


\begin{thebibliography}{99}
\bibitem{SimDet}
NVT simulations were carried out using Gromacs software  \cite{gromacs} using the Nos\'{e}-Hoover thermostat \cite{NoseHoover}. Molecules were kept rigid using the LINCS \cite{LINCS} algorithm.
\bibitem{Dumbbell} 
The large LJ sphere mimicking the phenyl group is similar to the one in the Lewis-Wahnstr\"om OTP model \cite{wahn94} with the parameters $m_p=77.106$ u, $\sigma_p=0.4963$ nm and $\epsilon_p=5.726$ kJ/mol. A small sphere mimicking the methyl group was taken from UA-OPLS \cite{Jorg84} with $m_m=15.035$ u, $\sigma_m=0.3910$ nm and $\epsilon_m=0.66944$ kJ/mol. 
Bond length: $d=0.29$ nm. The interaction between unlike particles is governed by the Lorentz-Berthelot mixing rules.
\bibitem{Equil} All state points were simulated for at least the time $t_{10}$, where $\left< \Delta r^2(t_{10}) \right> = 10$nm$^2$, i.e. on average the large spheres moved at least $6\sigma_p$ during the simulations. Equilibration runs were performed for at least the time $t_{10}$.
\bibitem{OTP}
$N=324$ molecules consisting of three LJ particles (with $\sigma=0.483$ nm, $\varepsilon= 600 $ K$k_B \simeq 4.989$ kJ/mol and $m=76.768$ u) placed in the corners of a rigid isosceles triangle with two sides of length $\sigma=0.483$ nm and one angle of $75^{\circ}$ \cite{wahn94}. LJ potentials were cut at $r_c=2.5\sigma$ using a shift function \cite{gromacs} with $r_1=2.3\sigma$. At the end of long simulations (0.1 $\mu$s to 5$\mu$s), the system crystallized at several state points (data not included in analysis): $T\leq400$ K at $\rho=$1.149 g/ml, $T\leq 400$ K at $\rho=$1.135 g/ml, $T\leq 350$ K at $\rho=$1.135 g/ml, and $T\leq 260$ K at $\rho=$1.093 g/ml ($\langle p\rangle=0$ kbar).
\bibitem{OTP-R}
At zero pressure we find $R<0.9$ for $T>400K$. Except for exact soft-spheres, the ``strongly correlating''-property holds only for a region of state points \cite{ped_prl}.
\end{thebibliography}
\end{document}